\documentclass[aps,prl,twocolumn,showpacs]{revtex4-1}
\usepackage{graphicx}

\begin{document}

\title{Dynamic facilitation observed near the colloidal glass transition}

\author{Scott V. Franklin}
\email{svfsps@rit.edu}
\affiliation{Department of Physics, Rochester Institute of
Technology, Rochester, NY, 14623-5603, USA}

\author{Eric R. Weeks}
\email{erweeks@emory.edu}
\affiliation{Department of Physics, Emory University, Atlanta,
Georgia 30322, USA}

\date{\today}

\begin{abstract}
We present experimental confirmation of dynamic facilitation in
monodisperse and bidisperse colloidal suspensions near the glass
transition volume fraction. Correlations in particle dynamics are seen
to exist not only in space (clusters and strings) but also as bubbles
in space-time. Quantitatively, highly mobile particles are more likely
(than immobile particles) to have nearest neighbors that were highly
mobile in immediately preceding times. The interpretation is that a
particle's mobility enables or facilitates the subsequent motion of
its neighbors. Facilitation is most pronounced at the relaxation time
that corresponds with cage-breaking, when dynamic heterogeneity is
also maximized.
\end{abstract}


\maketitle

As the temperature in a supercooled liquid is lowered, the dynamics
slow dramatically and, at a well-defined temperature, the system
becomes a glass\cite{cavagna09,ediger12,biroli13}.  The glass is
amorphous, with no long-range molecular ordering.  In liquid samples
close to the glass transition, molecular motion becomes slow and
spatially heterogeneous \cite{hempel00,sillescu99,ediger00}, occurring
through localized groups of molecules simultaneously rearranging
\cite{kob97,donati98}.  Approaching the glass transition, the
frequency of these motions drops and the length scales characterizing
the groups of rearranging particles increases
\cite{donati98,hempel00, hurley95}.

Garrahan and Chandler\cite{garrahan02,garrahan03} proposed a model
that posits the correlation of particle mobility not only in space but
also in space-time. In the {\it dynamic facilitation} model, motion by
a group of particles at one time facilitates {\it subsequent} motion
of particles in adjacent regions. Dynamic facilitation clusters,
``bubbles'' in space-time \cite{chandler10}, are distinct from
conventional clusters or strings of moving particles \cite{donati98},
which are spatially correlated only over a single time interval.
These space-time clusters can be quite large \cite{vollmayrlee06},
showing that local groups of mobile particles are more than just a
single-time random event. 

Vogel and Glotzer \cite{vogel04} found evidence for dynamic
facilitation in molecular dynamics simulations of viscous silica.
They observed that mobile particles, rather than randomly distributed
throughout the sample, were more likely to be found next to a
previously mobile particle.  They quantified this with a dynamic
facilitation parameter $F(\Delta t)$, defined as the increased
probability of a particle mobile over time $\Delta t$ having a
previously mobile nearest neighbor, as compared to the null hypothesis
of the previously mobile particles being randomly distributed.  In
their simulations, decreasing the temperature brought about the
development of a pronounced peak in $F(\Delta t)$ at a delay time
corresponding to the cage-rearrangement time scale. Quantitatively,
mobile particles were 1.5 to 2 times more likely than non-mobile
particles to have had at least one previously mobile nearest neighbor.
There has been, however, no direct experimental confirmation of
dynamic facilitation.

Colloidal suspensions provide a promising model system to test the
prediction of facilitated mobility.  Colloids are composed of
micron-sized solid particles in a liquid.  Our interest is in
colloidal glasses, similar to hard sphere glasses in that temperature
plays no role in the glass transition, its effects limited to particle
Brownian motion \cite{pusey86}.  Instead, the control parameter for
each sample is the particle volume fraction $\phi$.  Colloids have a
glass transition at $\phi_g \approx 0.58$, and are liquid-like at
volume fractions below this threshold \cite{pusey86,hunter12rpp}.  The
colloidal glass transition shares many similarities to the traditional
glass transition \cite{hunter12rpp}, such as a dramatically growing
viscosity \cite{cheng02}, increasing relaxation time scales
\cite{vanmegen94,brambilla09}, and amorphous structure similar to a
liquid \cite{vanblaaderen95}.  Most relevant for dynamic facilitation,
colloidal samples can be viewed directly with confocal microscopy:
experiments observed dynamic heterogeneity with long-range spatial
correlations \cite{kegel00,weeks00,narumi11}.  While these experiments
found growing dynamical length scales as the glass transition was
approached, the fundamental dynamics in question were all within a
specific time interval, probing simultaneous spatial correlations of
motion.  Testing the dynamic facilitation model, however, requires the
comparison of motion in two successive time intervals.

In this Letter we examine previously published data from confocal
experiments on colloidal samples close to the colloidal glass
transition \cite{weeks00,narumi11}.  These data sets are 3D
observations of the trajectories of several thousand particles per
sample.  We find dynamic facilitation occurs in these samples, with
the Vogel-Glotzer parameter reaching values similar to that seen in
simulations \cite{vogel04}.

{\it Experimental methods ---} We re-analyze data reported on
previously in Refs.~\cite{weeks00,narumi11}, which should be consulted
for complete experimental details. Both experiments involved colloidal
poly-(methylmethacrylate) particles sterically stabilized by a thin
layer of poly-12-hydroxystearic acid.  Weeks {\it et al.} studied a
nominally monodisperse sample particles with mean diameter
$d=2.36\ \mu$m \cite{weeks00}; Narumi {\it et al.} a nominally
bidisperse mixture of small ($d=2.36\ \mu$m) and large
($d=3.01\ \mu$m) particles with a number ratio of $N_S/N_L=1.56$
\cite{narumi11}.  All particles for both experiments were produced at
the University of Edinburgh by Andrew Schofield, and each particle
species had a polydispersity of 5\% \cite{weeks00,narumi11}.

We focus on data from liquid-like samples with $0.4 \leq \phi
\leq \phi_g$.  $\phi$ was determined from counting particles seen
in three-dimensional (3D) confocal microscope images, as described
below.  Uncertainties in the mean particle diameters result in
systematic uncertainties of the volume fraction \cite{poon12}
of $\pm 0.03$ for the monodisperse samples and $\pm 0.02$ for the
bidisperse samples.  As in the prior publications, we report volume
fractions to two significant digits (e.g., $\phi=0.55$), and they
are accurate to this extent relative to each other for a given
colloid type (monodisperse or bidisperse) \cite{poon12}.  
These particles are slightly charged, but nonetheless
the glass transition for each of these experimental data sets is
$\phi_g \approx 0.58$ \cite{weeks00,narumi11}.

Colloidal particles were imaged using confocal microscopy.  3D image
stacks were acquired on time scales sufficiently fast that particle
motion was minimal between each stack, which was straightforward given
the slow dynamics of the dense samples.  Particle
motion was tracked in 3D using standard techniques
\cite{crocker96,dinsmore01}.  For the monodisperse case, particle
positions are determined to $\pm 0.03\ \mu$m in $x$ and $y$, and $\pm
0.05\ \mu$m in $z$ \cite{weeks00,dinsmore01}.  For the bidisperse
case, the uncertainties were higher, $\pm 0.2\ \mu$m in $x$ and $y$,
and $\pm 0.3\ \mu$m in $z$ \cite{narumi11}.  This was due to the
difficulty of tracking both particle sizes simultaneously.

\begin{figure}
\includegraphics[width=8cm]{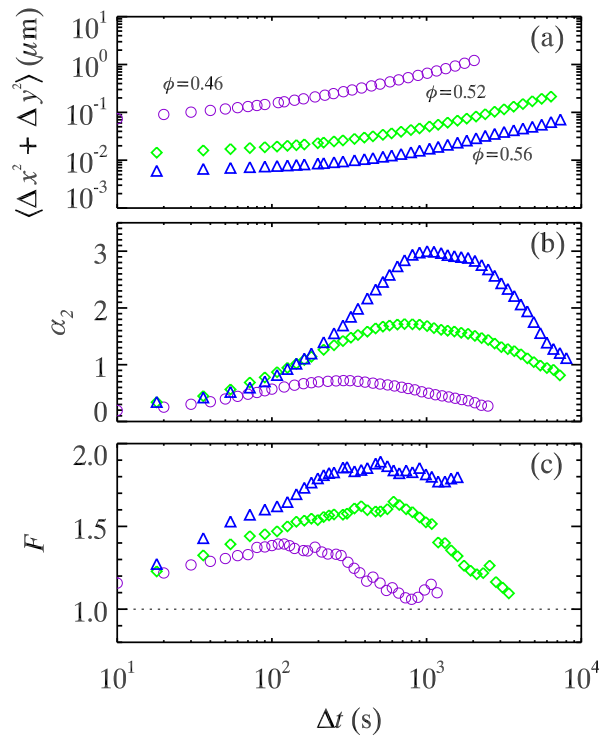}
\caption{\label{mono}(color online) (a) Mean-squared displacements
  (MSD) of three different volume fractions as indicated, of a
  monodisperse colloidal suspension.  (b) Non-Gaussian parameter and
  (c) dynamic facilitation parameter for the samples.  The symbols
  are the same in all three panels, defined as shown in (a).
}
\end{figure}

\begin{figure}
\includegraphics[width=8cm]{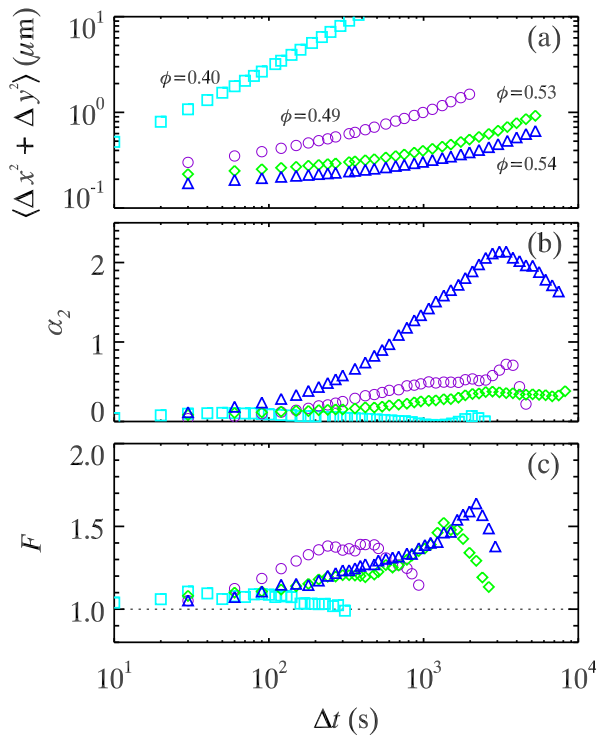}
\caption{\label{binary}(color online) (a) Mean-squared displacements
  (MSD), (b) non-Gaussian parameter and (c) dynamic facilitation
  parameter of small particles in a bidisperse colloidal suspension
  with volume fractions indicated in (a).  The symbols are the same in
  all three panels.  }
\end{figure}

The mean square displacement (MSD) gives a sense of the relevant time
scales and is shown in Figs.~\ref{mono}(a) and~\ref{binary}(a) for the
monodisperse and bidisperse samples respectively.  At time scales
shorter than shown ($\Delta t \lesssim 1$~s), particle motion is
diffusive.  At intermediate time scales the MSD's show a plateau that
develops with increasing volume fraction.  This represents
cage-trapping in which particle motion is constrained by nearest
neighbors.  At longer time scales, the cages rearrange and particles
move to new locations \cite{weeks00,narumi11}.  The rearrangement time
scale also corresponds to the time scale in which displacement
probability distributions are broader than a Gaussian.  This is
quantified by the non-Gaussian parameter $\alpha_2$, defined as
$\alpha_2(\Delta t) = (\langle \Delta x^4 \rangle/3\langle \Delta x^2
\rangle^2) -1$ \cite{kob97,rahman64}.  This parameter is zero for a
Gaussian distribution, and positive when the tails of the distribution
are more probable than expected for a Gaussian.  $\alpha_2(\Delta t)$
is shown in Figs.~\ref{mono}(b) and \ref{binary}(b).  Note that we
compute both the MSD and $\alpha_2$ using only the $x$ and $y$
displacements as they have less particle tracking uncertainty than the
$z$ component.  The non-Gaussian parameter $\alpha_2(\Delta t)$ peaks
at a time scale $\Delta t^\ast$, matching the end of the plateaus in
the MSD's.

\begin{figure}
\includegraphics[width=\columnwidth]{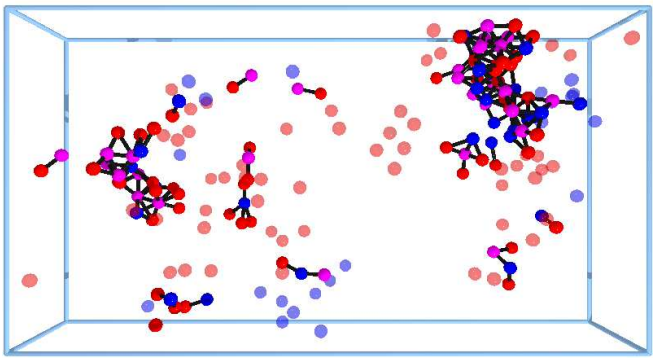}
\caption{\label{mclust}(Color) Highly mobile (see text) colloidal
  particles for two successive time intervals. Blue particles are
  those mobile in the first interval but not the second; red are those
  not mobile in the first interval but mobile in the second; and
  violet are those mobile during both intervals. Black bonds connect
  all particles mobile in the second interval to all neighboring
  particles mobile during the first interval. Particles not involved
  in facilitation (as measured by Eqn.~\ref{vgparam}) are not
  connected to other particles and drawn slightly transparent.  The
  particles are in their positions at the end of the first
  interval/beginning of the second interval.  Particles not mobile
  during either interval are not shown.  Particles are drawn at 60\%
  of their size.  The data correspond to a monodisperse sample with
  $\phi=0.52$, $\Delta t=600$~s. The displayed volume is $63\times
  34\times 11\ \mu$m$^3$.  }
\end{figure}

The larger particle displacements occurring on this time scale $\Delta
t^\ast$ involve spatially localized groups of particles
\cite{kob97,donati98,kegel00,weeks00,narumi11}.  Particles at a given
moment in time with unusually large displacements are typically
observed in clusters, as shown in Fig.~\ref{mclust} and discussed in
detail in Refs.~\cite{donati98,kegel00,weeks00,narumi11}.  All
particles, even those not rearranging on the time scale $\Delta
t^\ast$, are continually undergoing Brownian motion, and the
displacements of those that are truly rearranging are not much larger
than that of the caged particles \cite{weeks00}.  To best highlight
the mobile particles we consider a modified form of the displacement
vector first used by Donati {\it et al.}  \cite{donati98}:
\begin{equation} 
\label{displacement} 
\delta r
(t_0, \Delta t) = \max_{t_1,t_2} |\vec{r}(t_1) - \vec{r}(t_2)|
\end{equation} 
\noindent with $t_0 \leq t_1 < t_2 \leq t_0+\Delta t$.  This maximal
displacement removes effects due to random motions within a cage and
better captures the net particle displacement.  Using this definition
for mobility $\delta r$, we show the 5\% most mobile particles at two
moments in time in Fig.~\ref{mclust}.  The blue particles are mobile
in the first time interval, the red in the second, and purple in both.
Clusters of mobile particles are seen in several locations.  Red and
blue particles appear in similar locations; nearest neighbors (defined
below) are indicated by black bonds.  These are the particles for
which dynamic facilitation may play a role.

\begin{figure}
\includegraphics[width=7cm]{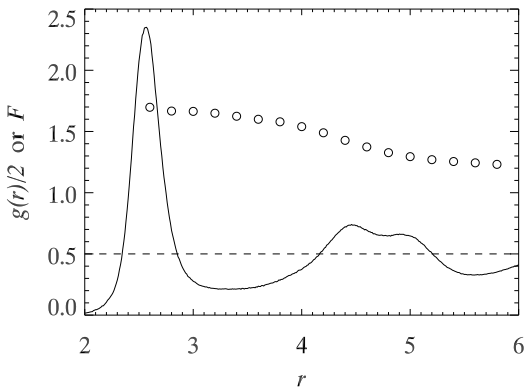}
\caption{\label{facnn}{\it Solid line:} Radial distribution function
  $g(r)$ (divided by 2 to fit within axes) for a monodisperse sample
  ($\phi=52\%$). The function peaks at the particle diameter
  2.36$~\mu$m. Nearest neighbors are defined as typically defined as
  all particles whose center-of-mass falls within the first minimum
  ($3.4\ \mu$m). {\it Circles}: Peak of dynamic facilitation parameter
  as a function of $r$, defining nearest neighbors as all
  particles within a distance $r$. For these data, the time scale
  is set to $\Delta t=600$~s, which maximizes $F$ [see diamond
  symbols in Fig.~\ref{mono}(c)].
  }
\end{figure}

{\it Dynamic Facilitation --- }Following Vogel and Glotzer, we define
two time intervals of duration $\Delta t$: the ``past'' interval
between $t-\Delta t$ and $t$ and the ``future'' interval between $t$
and $t+\Delta t$ and use Eqn.~\ref{displacement} to compute all
particles' mobilities in these two intervals.  We then define ``high
mobility'' particles as those with $\delta r(\Delta t)$ above a cutoff
$\delta r^*$ such that, over time, 5\% of the particles have
displacements larger than $\delta r^*$; although at any given moment,
the fraction of mobile particles may be more or less than 5\%
\cite{donati98,weeks00,vogel04}.  We are interested in particles with
both a low past mobility and a high future mobility, and compute the
probability $p_{LH}$ that these particles have at least one previously
high mobility neighbor.  The crux of the dynamic facilitation model is
that the transition in these particles from low-to-high mobility is
brought about by a previous high mobility neighbor.  (Nearest
neighbors are defined as all particles whose center-of-mass falls
within the first minimum $r_{min}$ of the radial distribution function
$g(r)$, shown by the solid line in Fig.~\ref{facnn} for a
representative sample.)  

As in Vogel and Glotzer, we also compute the probability $p_{LA}$ that
a particle with a low past mobility and {\it any} mobility in the
future has a neighbor with previously high mobility and define the
facilitation parameter as \cite{vogel04}
\begin{equation}
\label{vgparam}
F(\Delta t) \equiv p_{LH}/p_{LA}.
\end{equation}
$F$ is a function of the time interval $\Delta t$ used to define
mobility, and the probabilities are computed from all particles and
all times $t$.  If dynamic facilitation is not the mechanism for
correlated motion, then particles which become mobile (a low-to-high
mobility transition) bear no spatial relation to the previous high
mobility particles and $F(\Delta t)=1\ (p_{LH}=p_{LA})$.  Instead, as
we now describe, we find $F(\Delta t)>1\ (p_{LH} > p_{LA})$, and
conclude that dynamic facilitation {\it is} enabling large-scale
particle motions.  Conceptually, $F>1$ means that the particles
connected by black bonds in Fig.~\ref{mclust} are more prevalent than
expected by chance.

The resulting facilitation parameter as a function of delay time
$\Delta t$ is shown in Figs.~\ref{mono}(c) and \ref{binary}(c). For
both monodisperse and bidisperse samples the function shows a peak
that grows as the glass transition is approached.  The time scale
$\Delta t$ of this peak corresponds to that of the peak in the
non-Gaussian parameter. Figures~\ref{mono}(c) and \ref{binary}(c)
indicate that, compared to immobile particles, newly mobile particles
are almost twice as likely to have a previously mobile particle,
quantitatively (as well as qualitatively) similar to the simulations
results of Vogel \& Glotzer \cite{vogel04}.  Thus we find previously
mobile particles indeed facilitate the mobility of their neighbors as
predicted by Garrahan and Chandler \cite{garrahan02,garrahan03}.  As a
control case, the bidisperse sample with $\phi=0.40$ (far from the
glass transition) in Fig.~\ref{mono}(c) shows almost no facilitation.

To determine the spatial range over which facilitation occurs we vary
the distance $r$ used to define nearest neighbors.  The circles in
Fig.~\ref{facnn} show the peak value of facilitation $F$ as a function
$r$ (using $\Delta t$ close to the cage-rearrangement timescale
$\Delta t^*$).  Recall that the peak value of $F$ indicates the
increased probability that a newly mobile particle has at least one
previously mobile particle within the distance $r$.  The largest
values of $F$ are found for $r \leq 3.4$~$\mu$m, which matches the
conventional definition of a nearest neighbor.  Figure~\ref{facnn}
shows that this enhanced probability decreases with $r$, but is still
larger than 1 even over distances corresponding to second-nearest
neighbors (particles within the 2nd minimum of $g(r)$).  The decrease
for $r > 3.2$~$\mu$m shows that, in these colloidal samples,
facilitation is primarily through the influence of the nearest
neighbor particles.

In summary, we see spatiotemporal correlations of mobility in
colloidal samples approaching the glass transition that are the
defining feature of dynamic facilitation.  Quantitative evidence is
seen in a peak in the Vogel-Glotzer parameter close to the
rearrangement timescale $\Delta t^\ast$, the time scale for which
particle motion is spatially heterogeneous and cooperative
\cite{kob97,donati98,kegel00,weeks00}, and agrees with earlier
simulation results.  Our new analysis reveals that the spatially
correlated motion at a particular time facilitates subsequent motion
in adjacent particles, as predicted by the dynamic facilitation model.
Our data demonstrate that in samples close to the glass transition,
regions of mobility influence subsequent dynamics nearby, rather than
mobility arising randomly in new locations.  It is precisely these
facilitated spatially heterogeneous motions that eventually propagate
throughout the sample and allow it to flow, albeit on long time scales
for highly viscous samples near the glass transition.

\begin{acknowledgments}
This work was supported by the National Science Foundation through
grants DMR-1133722 (SVF) and CMMI-1250235 (ERW).
\end{acknowledgments}

\bibliography{eric}

\end{document}